\documentclass[prl,twocolumn,amsmath,amssymb,superscriptaddress,floatfix]{revtex4}

\usepackage{graphicx}
\usepackage{color}
\usepackage{soul}
\usepackage{epsfig}
\usepackage{caption}
\usepackage[normalem]{ulem}
\begin{document}
\title{Emergence and percolation of rigid domains during colloidal glass transition }
\date{\today}

\author{Xiunan Yang$^+$}
\affiliation{Beijing National Laboratory for Condensed Matter Physics and Key Laboratory of Soft Matter Physics, Institute of Physics, Chinese Academy of Sciences, Beijing 100190, People's Republic of China}
\affiliation{University of Chinese Academy of Sciences, Beijing 100049, People's Republic of China}
\author{Hua Tong$^+$}
\affiliation{Department of Fundamental Engineering, Institute of Industrial Science, University of Tokyo, 4-6-1 Komaba, Meguro-ku, Tokyo 153-8505, Japan}
\author{Wei-Hua Wang$^*$}
\affiliation{Institute of Physics, Chinese Academy of Sciences, Beijing 100190, People's Republic of China}
\affiliation{University of Chinese Academy of Sciences, Beijing 100049, People's Republic of China}
\affiliation{Songshan Lake Materials Laboratory , Dongguan, Guangdong 523808, China}
\author{Ke Chen$^*$}
\affiliation{Beijing National Laboratory for Condensed Matter Physics and Key Laboratory of Soft Matter Physics, Institute of Physics, Chinese Academy of Sciences, Beijing 100190, People's Republic of China}
\affiliation{University of Chinese Academy of Sciences, Beijing 100049, People's Republic of China}
\affiliation{Songshan Lake Materials Laboratory , Dongguan, Guangdong 523808, China}

\begin{abstract}

Using video microscopy, we measure local spatial constraints in disordered binary colloidal samples, ranging from dilute fluids to jammed glasses, and probe their spatial and temporal correlations to local dynamics during the glass transition. We observe the emergence of significant correlations between constraints and local dynamics within the Lindemann criterion, which coincides with the onset of glassy dynamics in supercooled liquids. Rigid domains in fluids are identified based on local constraints, and demonstrate a percolation transition near glass transition, accompanied by the emergence of dynamical heterogeneities. Our results show that the spatial constraints instead of the geometry of amorphous structures is the key that connects the complex spatial-temporal correlations in disordered materials.
\end{abstract}

\maketitle
A liquid solidifies when sufficiently cooled. Under near-equilibrium conditions, crystals form, with distinctively different structures and mechanical properties to the liquid phase. When rapidly quenched, on the other hand, a supercooled liquid undergoes glass transition and becomes an amorphous solid with apparently disordered structures. For the glass transitions, two fundamental questions remain. The first one is “is there a qualitative transition point between the liquid and solid phases during the glass transition?”. Glasses obviously fit our experiences with solids. Experimentally, however, there is no definitive signal for the emergence of rigidity, despite more than 10 orders of magnitude increase in viscosity during the glass transition. The other question is “what structural orders, if any, are associated with the unusual dynamical phenomena and the rise of rigidity during the glass transition?”. Many studies attempt to construct structural parameters based on local geometry to distinguish slow rigid domains from more mobile fluid regions in glasses~\cite{Spaepen1,Tanaka2,Ma1,Kelton,Hu}, but have yet to find any universal signatures. In condensed matters, particularly in solids, the role of the structure is to confine the motion of atoms, thus maintain rigidity. From this point of view, a solid lose its rigidity when the motions of consisting atoms can no longer be adequately constrained. A perfect example is the Lindemann criterion for the melting of crystals, which is found to be accurate in almost all crystalline materials~\cite{Lin1,Lin11}. A crystal melts when the vibrational fluctuations of atoms reach the order of 0.1 of the lattice constant. The Lindemann criterion is independent of the symmetry of the underlying structures of the solids, thus may be employed to determine the liquid-solid transition in glass forming materials~\cite{Lin2,Lin3,Lin4,Lin5,Lin6,Lin7,Lin8,Lin9,Lin10,Lin12,Lin13,Lin14,Lin15}. In meta-stable structures, the vibrational fluctuations of atoms are primarily determined by local structures, thus the confinement experienced by individual particles can be employed as a structural parameter when local geometry is too intricate to analyze. 

\begin{figure*}[t]
\includegraphics[width=18cm]{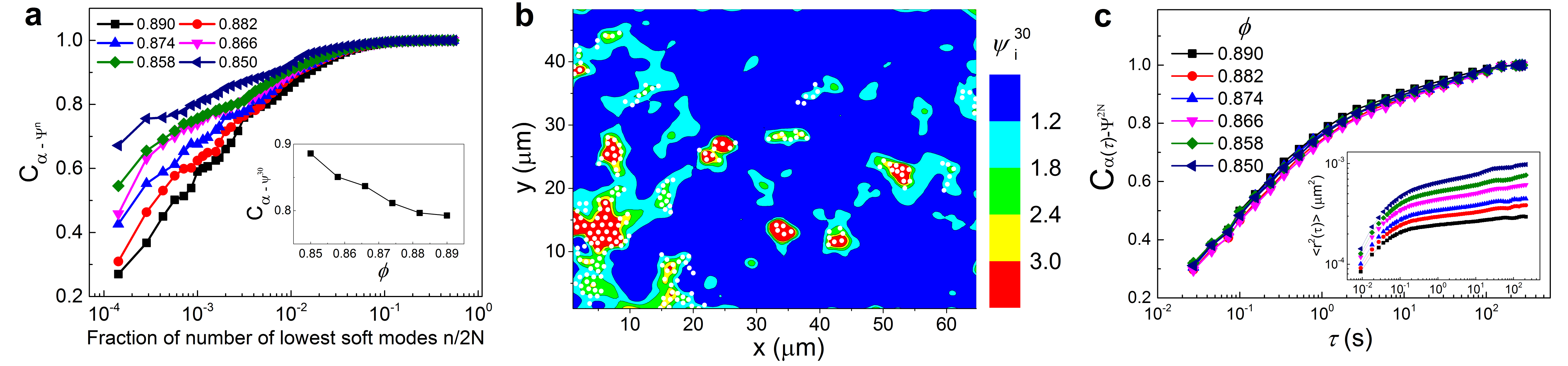}
\caption{\textbf{Correlation between soft mode and Debye-Waller factor in jammed packings.} {\bf a}, Correlation between $\alpha_{i}$ and $\Psi_{i}^{n}$ as a function of the fraction of the lowest frequency modes $n/2N$ included in $\Psi_{i}^{n}$ at different packing fractions. Inset: correlations between $\alpha$ and $\Psi^{30}$ at different packing fractions. The noise level is about 0.02. {\bf b}, Real space distribution of cooperatively rearranging regions (white dots) and $\Psi^{30}$ at $\phi=0.850$ (colored contours), normalized by the average value. {\bf c}, Rank correlations between $\Psi_i^{2N}$ and $\alpha_i(\tau)$ as a function of $\tau$ for different packing fractions. The noise level is about 0.02. Inset: MSDs at different $\phi$.}
\label{fig1}
\end{figure*}

In this Letter, we employ local Debye-Waller factor to measure the local constraints in colloidal liquids and glasses, and investigate its correlations to local dynamics during the glass transition. Temporal correlations between particle constraints and local dynamics reveal the emergence of structural relaxation barriers that give rise to finite rigidity in the system, as the temperature decreases. A common Lindemann-like length scale is identified by comparing the configurational changes when the system overcomes the relaxation barriers and starts behaving like fluids. The rise of rigidity and the onset of glassy dynamics are both shown to coincide with the percolation of rigid domains identified by the Lindemann-like length scale. Dynamical heterogeneity increases sharply when rigid domains percolate the system, and then decreases when the system becomes overwhelmingly solid. Our results suggest that a Lindemann-like criterion can be applied in amorphous materials to determine the transition between liquid and solid states, and the glass transition is the growth and percolation of rigid domains in supercooled liquids.

The samples consist of binary mixtures of poly-N-isopropylacrylamide (PNIPAM) particles~\cite{Yunker1,Chen2013} hermetically sealed between two coverslips, forming a monolayer of disordered packing. To avoid crystallization, the diameter ratio between large and small particles is chosen to be 1:1.4, with the number ratio close to 1. The PNIPAM particles are thermo-sensitive which allows the in-situ tuning of the packing fractions using an objective heater (BiOptechs). PNIPAM spheres are best described as hard spheres with soft shells~\cite{Yunker1,HanPRE}. At high packing fractions, PNIPAM particles are compressible to some extent, allowing observation of dynamical phenomena above the hard sphere jamming transition. The diameters of the particles are measured by dynamical light scattering to be 1 and 1.4 $\mu m$ at 22 $^{\circ}$C. The total number of particles in the field of view is about 3500. To cover a wide range of packing fractions, two groups of samples are seperately prepared. The packing fractions are between 0.890 to 0.850 (jammed solids) for the first group, and between 0.56 and 0.84 (unjammed liquids) for the second group. Here we use the 2D jamming packing fraction of hard spheres of 0.85 to indicate that no spontaneous topological rearrangements are observed in samples of higher packing fractions during the time window available to our experiments~\cite{Zhang1}. Before data acquisition, the samples are equilibriated on microscope stage for 3 hours. The particle configurations are recorded by digital video microscopy at 30 to 110 frames/s, and the particle trajectories are extracted by particle-tracking techniques~\cite{Grier1}. Combined optical and tracking error of particle fluctuations is estimated to be less than 0.01 $\mu m$ by measuring the MSD of fixed particles at different packing fractions. For jammed samples, the phonon modes are extracted using the covariance matrix analysis~\cite{Henkes1,Chen2010,Chen2011,suppl.}. The covariance matrix analysis measures the phonon modes of a “shadow system” with the same configurations and interactions as the colloids in experiment, but without the damping.

Spatial constraints felt by individual particles can be measured by either the lowest energy barrier for displacements or positional fluctuations. In jammed solids with stable configurations, the lowest energy barrier is directly related to the soft phonon modes~\cite{Xu1}. We employ a soft mode parameter $\Psi$ for individual particles, proposed by Tong and Xu~\cite{Xu2} based on equipartition hypothesis. For particle $i$, $\Psi_i^{2N} = \sum_{j=1}^{2N}\frac{1}{\omega_j^2}|{\vec{e}}_{j,i}|^2$, where $\omega_j$ is the vibrational frequency of mode $j$ and ${\vec{e}}_{j,i}$ is the polarization vector of particle $i$ in mode $j$, $N$ is the number of particles in a two-dimensional glass. $\Psi_i^{2N}$ is biased toward the lower frequency modes, as the contributions from high frequency modes to $\Psi_i^{2N}$ diminish rapidly with frequency. $\Psi_i^{2N}$ removes the ambiguities in soft modes selections, and can be proven to be statistically proportional to the single particle Debye-Waller factor $\alpha_i$ in meta-stable glasses (See supplementary for derivations~\cite{suppl.}). $\alpha_i= \langle[\vec{r}_i(t)-\vec{r}_i(0)]^2\rangle$, where $\vec{r}_i(t)$ is the position of particle $i$ at time $t$, and $\left< .\right>$ denotes the trajectory average~\cite{Xu2, Harrowell3}. Debye-Waller factor is often employed as a dynamical parameter. On short time scales when topological rearrangement is infrequent, local Debye-Waller factor is primarily determined by local structures, thus can be employed as a structural parameter as well. Previous experiments and simulations have shown that short-time local positional fluctuation is a good predictor of long-time dynamics in the supercooled and glass regime~\cite{Harrowell3,JPCL}.

The high correlations between soft mode $\Psi$ and Debye-Waller factor $\alpha$ are experimentally demonstrated in jammed colloidal glasses. Figure~\ref{fig1}a plots the Spearman's rank correlation between $\Psi_i^{n}$ and $\alpha_i$ as a function of the fraction of the lowest frequency modes $\frac{n}{2N}$ included in jammed colloidal glasses. The correlation to local dynamics comes predominantly from the lowest frequency modes, as the bottom $0.5\%$ of modes ($\sim30$ for our system) achieve a correlation over $0.8$. The inset of Figure~\ref{fig1}a plots the correlation between $\Psi_i^{30}$ and $\alpha_i$ at different packing fractions, which shows that in jammed solids, positional fluctuations of inidividual particles can be well decribed by a handful of soft modes. Figure~\ref{fig1}b shows the spatial distribution of cooperatively rearranging regions (CRRs) composed of the top $10\%$ fastest particles (white circles)~\cite{CRR1} and $\Psi_i^{30}$ (colored contours). It is clear that regions with higher concentration of soft modes are spatially correlated with fast local dynamics. 

\begin{figure}[t]
\includegraphics[width=8cm]{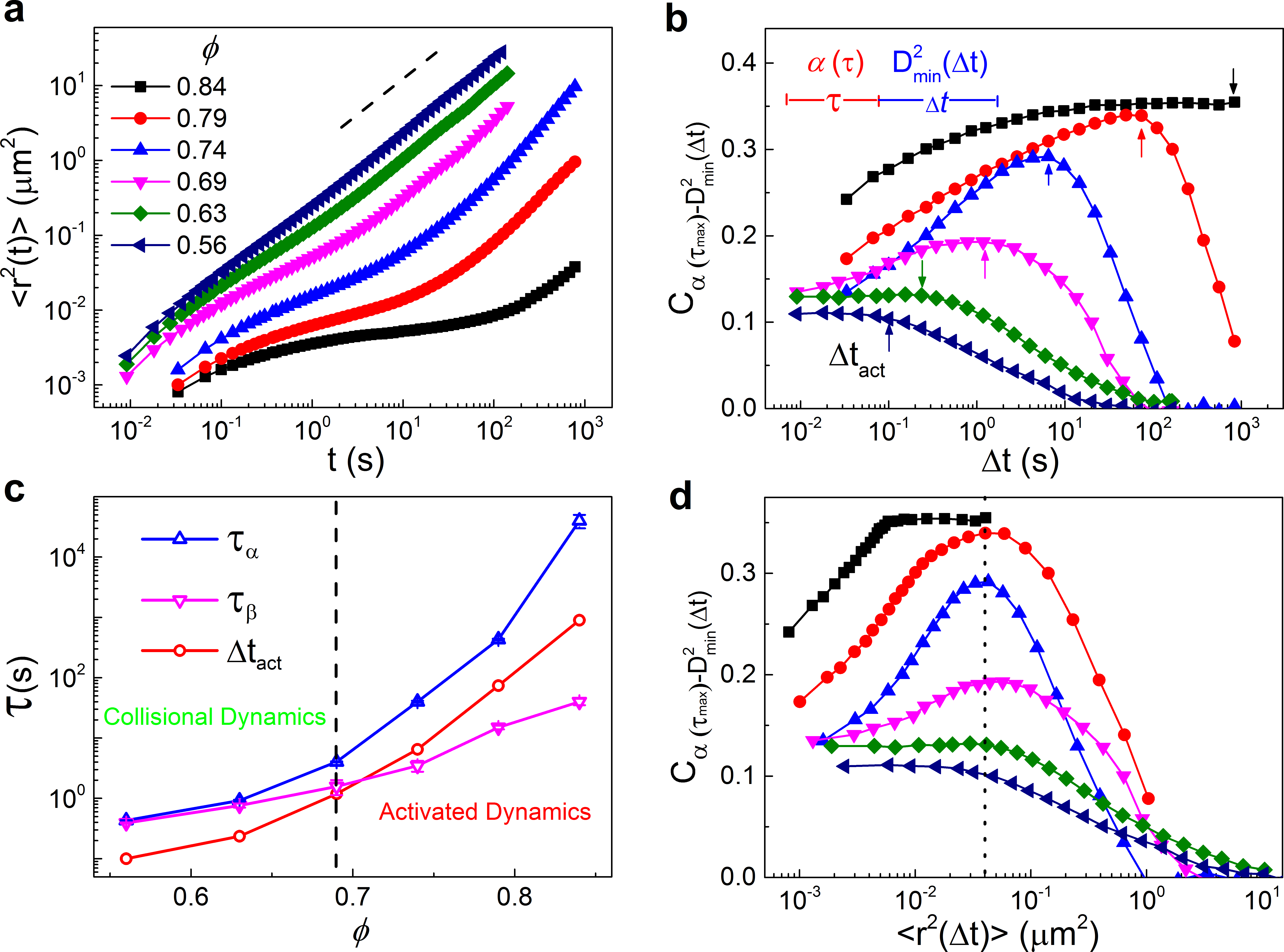}
\caption{\textbf{Structure-dynamics correlation during glass transition.} {\bf a}, Measured mean square displacements at different packing fractions. Dashed line indicates free diffusive motions.  {\bf b}, Spearman's rank correlation between $\alpha_i(\tau_{\rm max})$ and $D_{\rm min}^2(\Delta t)$ as a function of $\Delta t$. The vertical arrows indicate the $\Delta t_{\rm act}$ when the correlations start to decay. Inset: the time sequence for measuring $\alpha(\tau)$ and $D_{\rm min}^2(\Delta t)$. {\bf c}, $\phi$ dependence of the activation time $\Delta t_{\rm act}$, and the $\alpha$ and $\beta$ relaxation time. The dashed line indicates the onset of glassy dynamics. {\bf d}, MSDs dependence of correlation $C_{\alpha_i(\tau_{\rm max})-D_{\rm min}^2(\Delta t)}$ for each packing fraction. Vertical dotted line indicates the Lindemann criterion. The noise level is about 0.02.}
\label{fig2}
\end{figure}

In jammed glasses, soft modes can be accounted for by short-time fluctuations of particle positions. Figure~\ref{fig1}c plots the correlation between $\alpha_i(\tau)$ and $\Psi_i^{2N}$ as a function of the time window $\tau$ in which $\alpha_i$ is measured. The correlation increases rapidly for small $\tau$ values and reaches $\sim0.8$ at $\sim1 s$, within the $\beta$-relaxation time scale ($\sim10 s$) defined by the middle of the plateau in the log-log plot of the mean square displacements (Figure~\ref{fig1}c, inset)~\cite{Harrowell3}. The high correlations between short-time $\alpha_i(\tau)$ and $\Psi_i^{2N}$ suggest that the local structures can be adequately explored at relatively short periods of time. Further increasing of $\tau$ only slightly improves the correlation to soft modes. For comparison, it requires more than $1000 s$ of video microscopy measurements to properly extract the vibrational modes from the same jammed colloidal samples using covariance matrix analysis~\cite{Chen2010}. Thus short-time particle Debye-Waller factor can be employed as an effective soft mode parameter in colloidal systems below jamming~\cite{Royall2}, where direct measurements of spatial distribution of soft modes are difficult.

\begin{figure*}[t]
\includegraphics[width=14cm]{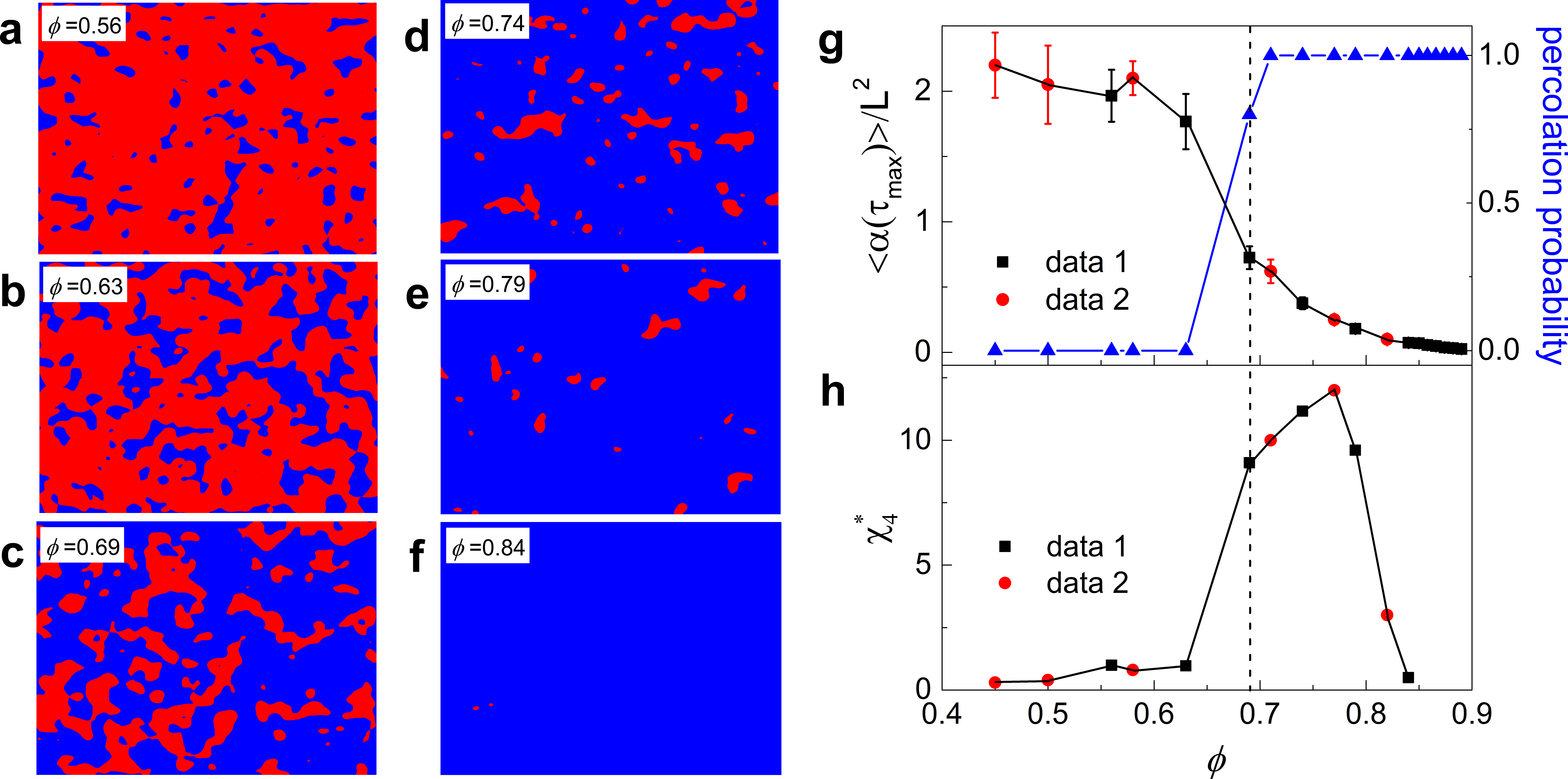}
\caption{\textbf{Structure evolution during glass transition.} {\bf a-f}, Spatial distribution of $\alpha_i(\tau_{\rm max})$ at different packing fractions, binarized by the Lindemann criterion. Red colors are fluid regions with $\alpha_i(\tau_{\rm max})$ larger than the Lindemann criterion; blue colors are rigid regions with $\alpha_i(\tau_{\rm max})$ below the Lindemann criterion. {\bf g}, Average $\alpha_i(\tau_{\rm max})$ and the percolation of rigid regions during glass transition. Left axis: average $\alpha_i(\tau_{\rm max})$ normalized by Lindemann criterion, as a function of $\phi$. The dashed line indicates the onset of glassy dynamics shown in Figure~\ref{fig3}c . The black squares (data 1) are measured from the same dataset as in Figure~\ref{fig1} and Figure~\ref{fig2}. To extend the range of the plot, we include measurements from an additional dataset (data 2, red circles). Error bars represent standard deviations. Right axis: The probability of rigid regions percolating the field of view(blue triangles). The probability is calculated as the fraction of the configurations with rigid regions percolating the field of view in all measured configurations. {\bf h}, Peak value of dynamical susceptibility, $\chi_4^*$ as a function of $\phi$.}
\label{fig3}
\end{figure*}

We now apply $\alpha_i(\tau)$ in unjammed colloidal liquids to measure local mechanical constraints.  The MSDs of the unjammed samples are plotted in Figure~\ref{fig2}a. As the Debye-Waller factor in unjammed fluids may vary with time, $\alpha_i(\tau)$ is no longer time averaged, instead it is calculated for each segment of trajectory in a time window of $\tau$. To identify the relevant time scales over which local structures have the most influence over future dynamics in liquids, we measure the temporal correlations between $\alpha_i(\tau)$ and local dynamics measured by non-affine displacement $D_{\rm min}^2(\Delta t)$~\cite{Yang,Langer} after the preceding structures are measured. $D^2(t_{1},t_{2})=\sum\limits_{n}\sum\limits_{i}[r^i_{n,t_{2}}-r^i_{0,t_{2}}-\sum\limits_{j}(\delta_{ij}+\varepsilon_{ij})\times(r^j_{n,t_{1}}-r^j_{0,t_{1}})]^2$
, where $r^i_{n,t}$ is the $i$th ($x$ or $y$) component of the position of the $n$th particle at time t, and the $\delta_{ij}+\varepsilon_{ij}$ that minimize $D^2$ are calculated based on $r^i_{n,t}$. $D^2_{\rm min}$ measures the particle level nonaffine strain, i.e., the minimum mean square difference between actual relative displacements of particle to its neighbors and the relative displacements that they would have if they were in a region of uniform strain. Correlations between $\alpha_i(\tau)$ and $D_{\rm min}^2(\Delta t)$ depend on both the window $\tau$ in which structural information is collected, and the timescale of the dynamics after $\alpha_i$ is measured, $\Delta t$. We choose the $\tau=\tau_{\rm max}$ that yields the highest correlations to $D_{\rm min}^2$~\cite{suppl.}. $\tau_{\rm max}$ is thus the proper time scale to identify structures that have the highest predictability for dynamics in liquids; and it naturally emerges from correlation measurements. For observation window shorter than $\tau_{\rm max}$, insufficient structural information is collected, and for much longer time windows, relevant information will eventually be lost in structural relaxations. In our experiments, $\tau_{\rm max}$ is found to be in the vicinity of $\beta$-relaxation time $\tau_\beta$~\cite{suppl.}, consistent with our results from jammed solids. The $\beta$-relaxation time and $\alpha$-relaxation time of the samples are extracted by fitting the intermediate scattering function with a two-step stretched exponential function (the Kohlrausch-Williams-Watts function)~\cite{Weitz1,Tanaka3,suppl.}. For liquids with only one-step relaxations, the fitting of the function yields two nearly identical relaxation times.

Figure~\ref{fig2}b plots the correlation between $\alpha_i(\tau_{\rm max})$ and $D_{\rm min}^2(\Delta t)$ as a function of $\Delta t$. The correlations are averaged over all available trajectories. At low packing fractions, the correlation between $\alpha_i(\tau_{\rm max})$ and local dynamics is low, and decays almost immediately after the $\alpha_i(\tau_{\rm max})$ is measured. This short memory in dynamics reflects a nearly flat potential energy landscape where structural relaxations are facilitated by free diffusion and collisions between particles. The energy landscape becomes more rugged as the packing fraction increases, and an activation mechanism begins to emerge~\cite{Lubchenko1}. At higher packing fractions, the correlation between $\alpha_i(\tau_{\rm max})$ and $D_{\rm min}^2$ first increases with $\Delta t$ then decreases after reaching a peak value at $\Delta t_{\rm act}$. This delayed correlation peak between local constraints and structural relaxations signifies the emergence of rearranging barriers, hence finite rigidity of the system, with $\Delta t_{\rm act}$ being the average time required for thermal fluctuations to overcome the barriers for structural relaxations. When the packing fraction is further increased, the energy barrier also increases, with higher peak correlation values.

The rise of the relaxation  barriers coincides with the separation of $\alpha$- and $\beta$-relaxation time scales in liquids~\cite{Debenedetti1}. Figure~\ref{fig2}c plots the measured $\tau_{\alpha}$ , $\tau_{\beta}$ and $\Delta t_{\rm act}$ in unjammed colloidal samples. The increase of the relaxation time is modest near the jamming point compared to standard hard sphere systems~\cite{SciRep,3D2}, due to the softness of PNIPAM spheres. Below $\phi=0.69$, $\Delta t_{\rm act}$ is short, and the $\tau_{\alpha}$ and $\tau_{\beta}$ are close. Without obvious peaks, $\Delta t_{\rm act}$ is chosen to be the  the point where the correlation between $\alpha_i(\tau)$ and $D_{\rm min}^2$ starts to decay, as indicated by vertical arrows in Figure~\ref{fig2}b. Around $\phi = 0.69$ where a delayed correlation peak appears, the $\alpha$- and $\beta$-relaxation times begin to separate. $\Delta t_{\rm act}$ becomes significantly larger than $\tau_{\beta}$ when the packing fraction is further increased. As the $\alpha_i(\tau)$  is measured on the time scale of $\tau_{\rm max}$ (close to $\tau_{\beta}$), a $\Delta t_{\rm act}$ greater than $\tau_{\beta}$ allows the prediction of long-time dynamics with short-time structural information. 

Temporally, local dynamics in liquids begin to decouple from earlier structures after $\Delta t_{\rm act}$. An interesting question is that do the average positional fluctuations of the particles reach a common length scale when the system begins to behave like a fluid, as in the case of the melting of crystals. In Figure~\ref{fig2}d, we replot the $C_{\alpha_i- D_{\rm min}^2}$ as a function of system MSDs. For all the packing fractions, the correlation begins to decay around $20\%$ of the small particle diameter $d$ indicated by the dashed line, close to the Lindemann criterion for the melting of crystals~\cite{Lin1}, despite orders of magnitude differences in relaxation time scales between these liquid samples. We can thus define $L = 0.2d$ as the equivalent melting criterion for glasses, and generalize the Lindemann criterion from the melting of crystals to the transition between solid and fluid phases in amorphous materials~\cite{Lin2,Lin3,Lin4,Lin5,Lin6,Lin7,Lin8,Lin9,Lin10,Lin11,Lin12,Lin13,Lin14,Lin15} where the dichotomy between solid and fluid phases has been ambiguous. For a given time window, structures that evolve less than the $L$ are considered solid-like or rigid, while structures evolve more than the $L$ are considered fluid-like.

Before applying the Lindemann-like criterion locally to identify rigid or fluidic domains, a proper observation time window needs to be determined. In the original Lindemann theory for crystals, the vibrational fluctuations of atoms around equilibrium positions are considered. For glasses, atoms can be considered primarily vibrating in cages on the $\beta$-relaxation time scale. However, instead of arbitrarily imposing the $\beta$-relaxation time, we employ the $\tau_{\rm max}$, which naturally emerges as the time scale most pertinent to future dynamics from inter-correlation measurement, as the observation window for the identification of rigid regions. Independent measurements confirm that the $\tau_{\rm max}$ in different samples are very close to the measured $\beta$-relaxation times~\cite{suppl.}. Using the time window of $\tau_{\rm max}$, we identify solid-like domains in unjammed samples whose $\alpha_i(\tau_{\rm max})$ are below the Lindemann criterion, and fluid regions with higher $\alpha_i(\tau_{\rm max})$ during the glass transition. Figure~\ref{fig3}a-f plot the snapshots of spatial distribution of $\alpha_i(\tau_{\rm max})$ at different packing fractions, binarized by the Lindemann criterion. A bond percolation based on the particle positions is used after we cluster rigid particles from the nearest neighbors which are determined from the first minimum of the radial distribution function. At low packing fractions, the system is mostly fluid-like (red color) with small pockets of solid-like regions (blue color). The rigid regions grow with the packing fraction and begin to percolate the system around $\phi=0.69$ until complete solidification near the jamming point. Key features of percolation phase transition are recovered by analyzing the distributions of the size and shape of the solid-like clusters~\cite{suppl.}. The percolation probability of rigid regions and the averaged $\alpha_i(\tau_{\rm max})$ of the system shows a sharp transition around $\phi=0.69$, as plotted in Figure~\ref{fig3}g.

The growth and percolation of the rigid regions in cooling liquids provide a microscopic origin for the onset of glassy dynamics shown in Figure~\ref{fig2}c and the dynamical heterogeneity. At low packing fractions (high temperatures), isolated rigid structures are created and relaxed by a one-step fluctuation-relaxation process. The size and the fraction of rigid regions both increase as the samples are further cooled. At a critical packing fraction ( $\phi=0.69$ in our experiments), the rigid regions become connected and percolate the system~\cite{Percolate1,Percolate2}. Before the percolation, isolated rigid domains exist in the liquid. However, unconnected rigid clusters cannot render the whole system rigid, as they are simply floating in a continuous phase of flowing liquid. Only after the percolation, the ability of the spanning network of rigid domains to resist small stresses gives rise to finite rigidity of the whole system [44, 45]. For the relaxation dynamics, before percolation, the rigid domains are formed and relaxed locally through fluctuations in the liquid, with a single relaxation time. After the percolation, while the liquid relaxation process remains in the liquid phase, the relaxation of the system-wide rigid network is much harder than isolated rigid clusters, which results in a much longer relaxation time, namely, the $\alpha$-relaxation time. The percolating rigid network also impedes long distance diffusions of particles. Under spatial confinement, particles are forced to rearrange locally through cooperative motions, or $\beta$-relaxation~\cite{CRR1,Length1,Length2}. The decoupling of relaxation times signals the transition from local relaxation process to a correlated relaxation process~\cite{JStatMech,CSA}. Dynamical heterogeneity naturally emerges from the competition between these two different relaxation mechanisms~\cite{DHpicture}. The peak of the dynamical susceptibility $\chi_4^*$ first increases around $\phi = 0.69$ and then decreases near the jamming point ($\phi_j\sim0.85$) when the whole system becomes homogeneously rigid~\cite{NP3}, as plotted in Figure~\ref{fig3}h (for the measurements of $\chi_{4}^*$, see the supplementary materials ~\cite{suppl., DH1, DH2}). 

In summary, by measuring the local constraints in colloidal liquids and glasses, we directly observe the emergence and growth of structure-dynamics correlations in supercooled liquids, which depend on a Lindemann-like length scale in configurational changes. The glass transition is then shown to be the growth and percolation of the rigid regions in supercooled liquids, which can be employed to explain the slowing-down and the dynamical heterogeneity~\cite{DHpicture,MCTpicture}. Although our results are obtained from a quasi-2D hard sphere colloidal system, the method to identify solid-like regions in fluids can be easily generalized to other glassy systems. Following the melting analogy, the rigid clusters in glass transition are similar to the crystalline nuclei during crystallization. But unlike the nuclei that span the system by growing from boundaries, the rigid clusters gain stability by forming a percolating network across the system. These clusters are also natural candidates for low-entropy droplets in random first-order transition theories for their slower dynamics~\cite{Lin3}. We thus speculate the percolation of rigid domains during glass transition can also be observed in 3D glasses~\cite{3D1,3D2,3D3,3D4,3D5} or in systems with different interactions, while the specific path leading to the percolation or the evolution of the connected rigid network after it may be different, which will be an interesting topic for future simulation or experimental studies. Our results are strong evidence that local constraints are a useful parameter to connect structure to dynamics in glassy systems compared to purely geometric or topological metrics. However, this discovery does not render the geometric structures irrelevant. It is obvious that the spatial constraints in glasses depend sensitively on local configurations, although specific dependence may vary greatly from system to system. It is only through the lens of the constraints can the correlations between structures and dynamics in disordered systems be clearly demonstrated. In addition, local constraints naturally include multi-body effects of amorphous structures that are difficult to quantify from analyzing the geometric structures alone. A direct link between conventional geometric structures and glassy dynamics may be established by searching for local and non-local configurations that contribute the most to local constraints in glassy materials~\cite{Andrea1}.\\

\textbf{Acknowledgements}
We thank Walter Kob, Peter Harrowell, Rui Liu, Mingcheng Yang, Chenhong Wang, and Maozhi Li for helpful discussions. This work was supported by the MOST 973 Program (No. 2015CB856800). K. C. also acknowledges the support from the NSFC (No. 11474327). \\
\end{document}